\documentclass{article}
\usepackage{arxiv}
\pdfoutput=1

\usepackage[utf8]{inputenc} 
\usepackage[T1]{fontenc}    
\usepackage{hyperref}       
\usepackage{url}            
\usepackage{booktabs}       
\usepackage{amsfonts}       
\usepackage{nicefrac}       
\usepackage{microtype}      
\usepackage{graphicx}
\usepackage{doi}
\usepackage{amsmath}
\usepackage{multirow}

\title{Modelling and forecasting patient recruitment in clinical trials with patients' dropout}

\date{}

\newcommand{\N}{\mathbb{N}}
\newcommand{\cL}{\mathcal{L}}
\def\a{\alpha}
\def\be{\beta}
\def\Ga{\Gamma}
\def\la{\lambda}
\def\si{\sigma}
\def\th{\theta}
\def\de{\delta}
\def\cB{{\mathcal B}}
\def\cD{{\mathcal D}}
\def\wti{\widetilde}
\def\what{\widehat}
\def\nn{\nonumber}
\def\d{\operatorname{d}}
\newcommand{\Pcro}[2][]{\mathbb{P}_{#1}\left[ #2 \right]}

\newcommand{\Esp}{\mathbb E\,}
\newcommand{\Pro}{\mathbb P\,}
\newcommand{\Var}{\mathbb V\,}
\newcommand{\Card}{\text{Card}}

\newcommand{\data}{\text{data}}
\def\Bin{{\rm Bin}}
\def\Gam{{\rm Ga}}
\def\Bet{{\rm Beta}}
\def\bee{\begin{equation}\label}
\def\ene{\end{equation}}
\def\beeq{\begin{eqnarray}\label}
\def\eneq{\end{eqnarray}}
\def\beeqn{\begin{eqnarray*}}
\def\eneqn{\end{eqnarray*}}
\newtheorem{remark}{Remark}

\author{
\and
{\large  Vladimir Anisimov$^1$},
 \
{\large Guillaume Mijoule$^2$}, \
{\large  Armando Turchetta$^3$},\
{\large     Nicolas Savy \thanks{E-mail: \texttt{Nicolas.Savy@math.univ-toulouse.fr}} $^{,2}$}
}

\renewcommand{\headeright}{}
\renewcommand{\undertitle}{}
\renewcommand{\shorttitle}{Modelling patient recruitment with patients' dropout}

\hypersetup{
pdftitle={Modelling and forecasting patient recruitment in clinical trials with patients' dropout},
pdfauthor={Nicolas Savy},
pdfkeywords={Patient recruitment, Poisson-gamma process, Prediction, Dropout},
}

\begin{document}
\maketitle

\vspace{-1.0cm}

\begin{center}
$^1$Data Science, Center for Design \& Analysis, Amgen, London, UK\\[2ex]

$^2$Institut de Math\'ematiques de Toulouse, Universit\'e de Toulouse, CNRS, UPS, UMR 5219, Toulouse, France\\[2ex]

$^3$Department of Epidemiology, Biostatistics and Occupational Health, McGill University, Montr\'eal, Canada\\[2ex]
\end{center}

\vspace{1.0cm}

\begin{abstract}
This paper focuses on statistical modelling and prediction of patient recruitment in clinical trials accounting for patients' dropout. The recruitment model is based on a Poisson-gamma model introduced by Anisimov \& Fedorov (2007), where the patients arrive at different centres according to Poisson processes with rates viewed as gamma-distributed random variables. Each patient can drop the study during some screening period. Managing the dropout process is of a major importance but data related to dropout are rarely correctly collected. In this paper, a few models of dropout are proposed. The technique for estimating parameters and predicting the number of recruited patients over time and the recruitment time is developed. Simulation results confirm the applicability of the technique and thus, the necessity to account for patient's dropout at the stage of forecasting recruitment in clinical trials.
\end{abstract}


\section{Introduction}

The problem of predicting patient recruitment and evaluating the recruitment time in clinical trials has been given much attention during the past years. However, till now some pharmaceutical companies still use techniques based on deterministic models and various \textit{ad hoc} techniques. Using a Poisson process to describe the  recruitment process is now an accepted approach (Senn, \cite{senn97,senn98}, Carter et al. \cite{cart-BMC05,cart-CCT04}). However, in real trials, the recruitment rates  in different centres vary and to mimic this variation it is natural to use a gamma distribution. Note that the use of Poisson-gamma mixtures for describing the variation of positive variables in modelling flows of various events has a long history, see e.g. Bates \cite{bates}.

For modelling patient recruitment  Anisimov and Fedorov \cite{an-fed-chapter-07,an-fed07} proposed to use a doubly stochastic Poisson process to take into consideration the variation in recruitment rates between different centres. This model, called a Poisson-gamma model, assumes that the patients arrive at different centres according to Poisson processes with the rates viewed as independent gamma distributed random variables. In Anisimov and Fedorov \cite{an-fed07} the procedure of parameters estimation at interim stage and the technique for predicting future recruitment process using empirical Bayesian technique have been suggested. The model has been validated using data from a large number of real trials \cite{an-outs09,an-fed07,ADF07}. This model was developed further for predicting recruitment process at the initial and interim stages to account for the situations when the centres opening dates may not be known and assumed to be  uniformly distributed in some intervals \cite{an-JSM09,ADF07,MS12}, extended also to using gamma and beta distributions for centres opening dates in \cite{A2020}, to the case where some centres can be closed or opened in the future \cite{an-comm-stat11}, and for sensitivity analysis to parameter errors \cite{MS12}. Some problems of optimal recruitment design accounting for time/cost constraints were considered in \cite{A2020}. This model was also used as a basis for developing techniques for the analysis of the effects of unstratified and centre-stratified randomization, predictive event modelling, and predicting randomization process \cite{an-comm-stat11}. Note that Gajewski et al \cite{GSC08} also modelled patient recruitment using exponential inter-arrival times with gamma-distributed parameter in Bayesian setting  but considered only the case of a trial with one clinical centre. Other approaches to recruitment modelling primarily deal with global recruitment. These approaches use different techniques, and we refer interested readers to survey papers by Barnard et al. \cite{pmid20604946}; Heitjan et al. \cite{pmid26188165}) and Gkioni et al. \cite{pmid31299358}, and also to a discussion paper Anisimov \cite{an-CCT-16} on using Poisson models with random parameters for patient recruitment modelling with other references therein.

Here, we use a Poisson-gamma model as a starting point for the patient arrival process and develop technique further assuming that each patient can be lost during the following screening process. Suppose that the screening interval, which is the time that a patient has to complete some preliminary tests for inclusion-exclusion criteria and to be randomized into the study, is a fixed positive number $R$ which is the same for all patients.  As the patients may fail to some tests, we assume that a patient can be lost either at the start of the screening process with some probability or during the screening interval at some random time. Although the collection of selection and recruitment data is recommended \cite{CONSORT}, in practice these data are rarely collected or at least not with adequate precision. Indeed, these data are of no practical use except to calibrate models like those proposed in this article.

The paper is organised as follows. In Section \ref{model}, we define the model. In Section \ref{estimation} the technique for estimating parameters
at the interim stage is provided, whereas Section \ref{prediction} is devoted to the prediction of the recruitment time using the parameters estimated in Section \ref{estimation}. Technical considerations concerning these sections are moved to the Appendix. Section \ref{simulation} illustrates these results by  simulation studies. The conclusions are provided in Section \ref{conclu}.

\section{Models for recruitment with patients' dropout.}\label{model}

\begin{figure}
\centering
\includegraphics[height=5cm]{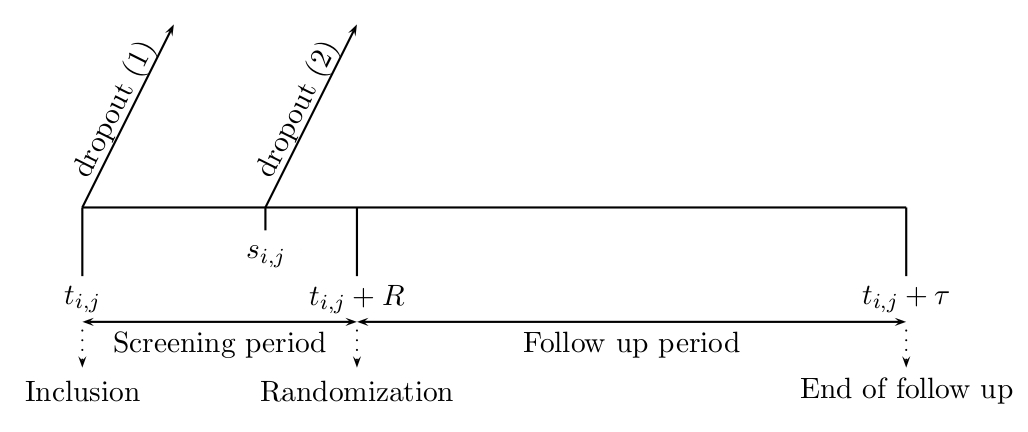}
\caption{Illustration of the dropout process for patients during screening.}
\label{F-DOF}
\end{figure}

Consider a multicentre study with $M$ clinical centres. Denote by $u_i$ the opening date of centre $i$. The patients arrive at centres according to independent doubly stochastic Poisson processes $\left\{N_t^i, t \ge 0 \; ; \; 1\leq i \leq M \right\}$
with time-dependent rates of the form $\la_i(t)=\la_i \mathbf{1}_{t \ge u_i}$. The values $\left\{\la_i \; ; \; 1\leq i \leq M\right\}$ are independent identically distributed random variables (i.i.d.r.v. for short) having a gamma distribution with some unknown parameters $(\a,\be)$. Denote by Ga$(\alpha,\beta)$ a gamma-distributed random variable with parameters $(\a,\be)$ and probability density function
\
$
p_{\a,\be}(x) = \be^\a \Gamma(\a)^{-1}e^{-\be x} x ^{\a-1}, \ x\geq 0.
$
Let $N_t = \sum_{i=1}^M N_t^i$ be the total number of patients arrived at all the centres in time interval $[0,t]$ and let $\{t_j, j\geq 1 \}$ be the increasing series of the jump times of $N_t$ (respectively, $\{t_{i,j}, j\geq 1 \}$ - jump times for $N_t^i$ in centre $i$, $ 1\leq i \leq M$). Consider modelling of the dropout effect. Patient may drop out the study
\begin{itemize}
\item at the time upon arrival,
\item during a fix screening period.
\end{itemize}
If neither one of these events happen, the patient is successfully randomized at time $s+R$, where $s$ is the arrival time, and registered to participate in the trial see Figure \ref{F-DOF}. Models for both these cases are proposed in the following sections.

\subsection{Dropout at the time upon arrival.} \label{modelDOUA}

Let us introduce the independent families of the i.i.d.r.v. $\{r_i \; ; \;  1 \le i \le M \}$ with values in $[0,1]$. Here $1-r_i $ stands for the probability of dropout upon arrival. Randomness in $\{r_i \}$ reflects the variation in these values across different centres.

Consider centre $i$ at some interim time $t_1$ and assume for simplicity that $t_1 > u_i$ where $u_i$ is the time of centre activation. Suppose that the values $r_i$ are given. Then for patient $j$ that has arrived at time $s=t_{i,j} < t_1$ there can be two events:
\begin{itemize}
\item  the patient is successfully randomized at time $s$ with probability
$$
p_i(s,t_1,r_i) = r_i;
$$
\item  the patient is lost with probability
$$
q_i(s,t_1,r_i) = 1-r_i.
$$
\end{itemize}
Let us define the independent families of indicators $\{\chi_{ij}(r_i), j \ge 1 \; ; \;  1 \le i \le M \}$, where for a given $r_i$ the variables $\{\chi_{ij}(r_i), j \ge 1\}$ are conditionally independent and for any $1 \le i \le M$ and any $j \geq 1$,
$$
\Pro (\chi_{ij}(r_i) = 0)= 1-\Pro (\chi_{ij}(r_i) = 1)=r_i.
$$
This means, if $\chi_{ij}(r_i)=0$, the patient $j$ in centre $i$ doesn't drop the study upon arrival.

Denote by $\Card(A)$ the number of points in the set $A$. Now, for each centre $i$, at any time $t \ge 0$, we define two processes:
\begin{itemize}
\item \textbf{randomized patients:}
$$
N_t^{i,R} = \Card\left\{ j \; : \; u_i\leq t_{i,j} \leq t \text{ and } \chi_{ij}(r_i) =0 \right\},
$$
\item \textbf{lost patients:}
$$
N_t^{i,L} = \Card\left\{ j \; : \; u_i\leq t_{i,j} \leq t \text{ and } \chi_{ij}(r_i) =1 \right\}.
$$
\end{itemize}
Finally, denote
$$
N_t^X = \sum_{i=1}^M N_t^{i,X} \qquad \text{for } X := R, L.
$$
The trial stops as soon as the desired number of randomized patients $N_R$ is reached, that is when $N_t^R \ge N_R$ - sample size.
We consider several models for dropout.

\noindent
\textbf{Model A.1. } For all $1 \le i \le M$, $r_i = r$ where $r$ is a fixed constant in $[0,1]$.\\[1ex]
\textbf{Model A.2. } The variables $\{r_i\; ; \;  1 \le  i \le M	 \}$ are i.i.d.r.v. having a beta distribution with  parameters $(\psi_1, \psi_2)$. This means, the variation in probability of randomization between different centres is described using a beta distribution.

\subsection{Dropout during screening.} \label{modelDODS} 

The model introduced in Section \ref{modelDOUA} is enriched by introducing the parametric family of positive random variables $\{Z_{ij}(\th_i), j \ge 1 \; ; \;  1 \le i \le M \}$, where for each $i$ and fixed $\th_i$ the variables $\{Z_{ij}(\th_i), j \ge 1\}$ have the same distribution, and the values $\{\th_i \; ; \;  1 \le i \le M \}$ are i.i.d.r.v. with some distribution. Here $1-r_i $ stands for the probability of  dropout upon arrival and $Z_{ij}(\th_i)$ - for time of dropout in center $i$. Randomness in $\{r_i \}$ and $\{\th_i \}$ reflects the variation in these values across different centres.

The patient $j$ arriving at centre $i$ at time $s=t_{i,j}$ may drop the study upon arrival with probability $1-r_i$ due to different initial tests. Otherwise, the patient may drop the study at some random time $s+Z_{ij}(\th_i)$ during the screening interval if $Z_{ij}(\th_i) \leq R$. 

Consider centre $i$ at some interim time $t_1$ and assume for simplicity that $R < t_1-u_i$. Suppose that the values $(r_i,\th_i)$ are given. Then at time $t_1$ for patient $j$ that has arrived at time $s=t_{i,j}<t_1$ there can be three events:
\begin{itemize}
\item  patient is successfully screened and randomized at time $s+R$ with probability
$$
p_i(s,t_1,r_i,\th_i) = r_i\Pro (Z_{ij}(\th_i) > R) \mathbf{1}_{s \le t_1-R};
$$
\item  patient is lost with probability
$$
q_i(s,t_1,r_i,\th_i) = 1-r_i+r_i\Pro (Z_{ij}(\th_i) \le \min(R, t_1-s) );
$$
\item patient is still in screening process with probability
$$
g_i(s,t_1,r_i,\th_i) = r_i\Pro (Z_{ij}(\th_i) > t_1-s)  \mathbf{1}_{t_1-R \le s <t_1}.
$$
\end{itemize}
Now, for each centre $i$, at any time $t \ge 0$, we define three processes:
\begin{itemize}
\item \textbf{randomized patients:}
$$
N_t^{i,R} = \Card\left\{ j \; : \; u_i\leq t_{i,j} \leq t- R \text{ and } \chi_{ij}(r_i) =0, \ Z_{ij}(\th_i) \geq R\right\},
$$
\item \textbf{lost patients:}
$$
N_t^{i,L} = \Card\left\{ j \; : \; u_i\leq t_{i,j} \leq t \text{ and } \{\chi_{ij}(r_i) =1\} \cup \{Z_{ij}(\th_i) \leq \min(R,t-t_{i,j})\} \right\},
$$
\item \textbf{patients in screening process:}
$$
N_t^{i,S} = N_t^i-N_t^{i,R}-N_t^{i,L}.
$$
\end{itemize}
Finally, denote
$$
N_t^X = \sum_{i=1}^M N_t^{i,X} \qquad \text{for } X := R, L, S.
$$
The trial stops as soon as the desired number of randomized patients $N_R$ is reached, that is when $N_t^R \ge N_R$ - sample size.
We consider several models for dropout.

\noindent
\textbf{Model B.1. } For all $1 \le i \le M$, $r_i = r$ and the values $\{Z_{ij}(\cdot), j \ge 1\; ; \;  1 \le  i \le M	\}$ are i.i.d.r.v. having an exponential distribution with parameter $\th$ (the same for all centres).\\[1ex]
\textbf{Model B.2. } For all $1 \le i \le M$, $r_i = r$ and the values $\{Z_{ij}(\cdot), j \ge 1\}$ given $\th_i$ are i.i.d.r.v. having an exponential distribution with parameter $\th_i$, where the values $\{\th_i\; ; \; 1 \le i \le M\}$ are i.i.d.r.v. having a gamma distribution with  parameters $(\a_2,\be_2)$.\\[1ex]
\textbf{Model B.3.} The variables $\{r_i\; ; \; 1 \le i \le M \}$ are i.i.d.r.v. having a beta distribution with some parameters $(\psi_1, \psi_2)$. The values $\{Z_{ij}(\cdot), j \ge 1\; ; \;  1 \le  i \le M	\}$ given $\th_i$ are i.i.d.r.v. having an exponential distribution with parameter $\th_i$, where the values $\{\th_i\; ; \; 1 \le i \le M\}$ are i.i.d.r.v. having a gamma distribution with  parameters $(\a_2,\be_2)$.\\

As we see, model B.3 is the most advanced model that accounts for the variation in the probability of dropout upon patient arrival and in the distribution of dropout time during screening process across clinical centres. For each model we consider the procedure of estimating unknown parameters and predicting in time the future process of randomized patients and the total recruitment time.

In models A.1 and A.2, the actual time of patient dropout during screening process is not taken into account. Thus, in the estimation procedure at some interim time $t_1$,  it is enough to know for each centre $i$ the number of recruited and randomized patients (that is, $\{ N_{t_1}^i, N_{t_1}^{i,R}\}$). On the other hand, in models B.1, B.2 and B.3, we assume that full data is available: for each patient  it is known the arrival and dropout (or randomization) time. If the dropout times were unknown, it would be impossible to distinguish between a patient lost upon arrival or during screening process, and the distinction within the model would be irrelevant.

\section{Parameters' estimation at interim time.} \label{estimation}

We use a Poisson-gamma recruitment model for modelling patients recruitment process \cite{an-fed07,an-world-congr08,an-JSM09,an-comm-stat11}. That means, recruitment rates $\la_i$ are viewed as a sample from a gamma distributed population with some unknown parameters $(\a,\be)$.

Let $t_1$ be some interim time. Consider some centre $i$
and assume for simplicity that $\tau_i=t_1-u_i \ge R$ (with $R=0$ for models A.1-A.2). In this case the number of recruited patients
at time $t_1$, $n_i=N_{t_1}^i$, as a random variable, has a negative binomial distribution with parameters $(\a,\tau_i/(\be+\tau_i))$ \cite[p.199]{john93} (or a Poisson-gamma distribution with parameters $(\a,\be,\tau_i)$ \cite[p.119]{bern04}). Recall that the probability distribution for a negative binomial distribution (Poisson-gamma) with parameters $(\a,\pi)$ is:
\begin{equation*}
{\rm NegBin}(k;\a,\pi) = \frac{\Gamma(\a+k)}{k!\Gamma(\a)}\pi^k(1-\pi)^\a.
\end{equation*}
Denote by $\mu = \Esp[\la]$ the mean recruitment rate and by $\si^2=\Var[\la]$ the variance of the rate and notice a useful relation between parameters: $\mu=\a/\be$, $\si^2=\a/\be^2$ which leads to $\a=\mu^2/\si^2$, $\be=\mu/\si^2$. Then, using the parametrization in terms of the mean rate  in centre~$i$,
\begin{equation} \label{negb1}
\Pro( n_i = k ) = \Esp\left[ e^{-\la \tau_i} \frac{(\la\tau_i)^{k}}{k!} \right] ={\rm NegBin}\left(k;\a, \frac{\mu\tau_i}{\a+\mu\tau_i}\right).\\
\end{equation}

\subsection{Dropout at the time upon arrival.} \label{S-DOUA}
\noindent
\textbf{Model A.1.}
Let $r_i$ be the probability of randomization in centre $i$. Denote by $\Bin(n,\pi)$ a binomial random variable  with parameters $(n,\pi)$ whose probability distribution is:
$$
{\rm Bin}(k;n,\pi) = {n \choose k} \pi^k (1-\pi)^{n-k}, \quad 0 \le k \le n.
$$
Assume for simplicity that there is no screening delay. Then in centre $i$ the number of  randomized patients $k_i$ has a binomial distribution with parameters $(n_i,r_i)$. Suppose that $r_i \equiv r$ (probability of randomization is the same). Then, given data in $M$ centres,
the log-likelihood function can be easily computed (see Appendix Section \ref{PARAMEST}) and yields to the maximum likelihood estimator
\begin{equation} \label{es32}
\what r = \Big(\sum_{i=1}^M n_i \Big)^{-1} \sum_{i=1}^M k_i.
\end{equation}
Note that if there is a screening delay, then such patients that entered screening process but the results of their screening procedure are unknown yet should be excluded in the calculations of probability of randomization, otherwise this probability will be underestimated. Therefore, instead of $n_i$ we should count $\wti n_i$, the number of patients with known screening results.

\begin{remark}
The independence of the dropout's and recruitment's processes implies that, whatever the model is, the parameters $(\a,\mu)$  can be estimated using a log-likelihood function $\cL_{1,1}$ given by relation \eqref{es3} in Appendix and a two-dimensional optimization procedure.
\end{remark}

\begin{remark}
Note that the estimator of the variance is $\what {\si}^2 = \what \mu^2/\what \a$.
\end{remark}

\textbf{Model A.2.}
Assume now that $r_i$ can vary between different centres and we describe this variation using a beta distribution with some unknown parameters $(\psi_1, \psi_2)$. Denote by $\Bet(\psi_1, \psi_2)$ a beta-distributed random variable with p.d.f.
$$
p_{\be}(x; \psi_1, \psi_2) = x^{\psi_1-1} (1-x)^{\psi_2-1}/\cB(\psi_1, \psi_2), \quad x \in ]0,1[,
$$
where $\cB(\psi_1, \psi_2)=\int_0^1 x^{\psi_1-1} (1-x)^{\psi_2-1} \d x$ is a beta function.\\
Notice the fact that if $r = \Bet(\psi_1, \psi_2)$, then a doubly stochastic binomial variable $\Bin(n, r)$ has a beta-binomial distribution whose probability distribution is:
$$
\Pro (\Bin(n; \Bet(\psi_1, \psi_2)) = k ) = {n \choose k} \frac {\cB(k+\psi_1, n-k+\psi_2)} {\cB(\psi_1, \psi_2)}, ~k=0,\dots,n.
$$
Parameters $(\psi_1,\psi_2)$  can be estimated using the log-likelihood function $\cL_{2,2}$ given by \eqref{eq6} in Appendix Section \ref{PARAMEST} and a two-dimensional optimization procedure. Denote the estimators of parameters by $(\what \a, \what \mu)$ and $(\what \psi_1, \what \psi_2)$. Consider now a Bayesian procedure of adjusting (re-estimating) parameters in each centre given data $n_i$ and $k_i$ in this centre similar to the one developped by Anisimov \cite{an-fed07,an-comm-stat11}. As $\la_i$ has a prior gamma distribution with parameters $(\a,\be)$, given data $\{n_i, \tau_i \}$ and using the Bayesian formula, one can calculate that the posterior distribution of $\la_i$ is also a gamma distribution with parameters $(\a+n_i, \be+\tau_i)$. Correspondingly, if $r_i$ has a prior beta distribution with parameters $(\psi_1,\psi_2)$, then, given data $\{n_i, k_i \}$, one can calculate that the posterior distribution of $r_i$ is also a beta distribution with parameters $(\psi_1+k_i, \psi_2+n_i-k_i)$ (see Bernardo et al. \cite{bern04} p. 267 for details). Therefore, given data, we can represent the posterior estimators of the rates and the probabilities of randomization in each centre in the form:
\begin{equation} \label{eg7}
\what \la_i = \Gam(\what \a + n_i, \what \be + \tau_i) \quad \text{and} \quad \what r_i = \Bet(\what \psi_1+k_i, \what \psi_2+n_i-k_i),
\end{equation}
where, given data, $\what \la_i$ and $\what r_i, i=1,..,M$ are independent.

\subsection{Dropout during screening.} \label{S-DODS}

For calculation of the likelihood function we need to account for that the variables $\{\la_i, r_i, \th_i\; ; \;  1 \le  i \le M	\}$ are independent and, in general, are some random variables. As we assume that the rates $\{\la_i \; ; \;  1 \le  i \le M	 \}$ are viewed as i.i.d.r.v. having a gamma distribution with parameters $(\a,\be)$, then for any $1 \le  i \le M	$ the variable $N_t^i$  has a negative binomial distribution ${\rm NegBin}\left(k;\a, \frac{\mu\tau_i}{\a+\mu\tau_i}\right)$ (see \eqref{negb1}). The types of distributions of ${r_i}$ and ${\th_i}$ are specified by the types of models B.1-B.3. In these models, we assume that more information is available, so  we can estimate parameters of  dropout times $\{ Z_{ij}(\theta), j \ge 1\; ; \;  1 \le  i \le M	\}$. For a sake of simplicity, in the sequel $Z_{ij}(\theta)$ will be denoted $Z_{ij}$. Moreover, the calculation of the posterior distributions of parameters is not straightforward if we use data as in models A. At time $t_1$, we observe patients arrival times $\{t_{i,j}\leq t_1\; ;\; j \ge 1\; , \;  1 \le  i \le M \}$ and the last times $\{s_{i,j}, j \ge 1\; ; \;  1 \le  i \le M	\}$  they were in the screening process (i.e $t_{i,j} \le s_{i,j} \le t_1$) . This means we also observe $\{ \min(Z_{ij}, R, t_1-t_{i,j})\vee 0, j \ge 1\; ; \;  1 \le  i \le M	\}$, where $a\vee b = \max(a,b)$.

Denote by $l_i$ the number of patients lost during the screening process in centre $i$,  by $T_i= \sum_j (s_{i,j}-t_{i,j})$  a sum of screening durations in centre $i$, and by $\tilde k_i$ the number of patients that are not lost immediately upon arrival.

\textbf{Model B.1.}
The maximum likelihood estimators  of $r$ and $\theta$ are given by
\begin{equation} \label{es33}
\what r = \Big(\sum_{i=1}^M n_i \Big)^{-1} \sum_{i=1}^M \tilde k_i \quad\text{and}\quad \what \th = \Big(\sum_{i=1}^M T_i \Big)^{-1} \sum_{i=1}^M l_i,
\end{equation}
details are given in Appendix Section \ref{PARAMEST}.

\textbf{Model B.2.}
Parameters $(\a_2, \beta_2)$  can be estimated using log-likelihood function $\cL_{4,2}$ given by \eqref{es34} in Appendix Section \ref{PARAMEST} and a two-dimensional optimization procedure however $r$ is estimated
as in \eqref{es33}.

Consider the Bayesian procedure of re-estimating parameters in each centre given data $\{n_i,\tau_i,(t_{i,j}),(s_{i,j}), j \ge 1\; ; \;  1 \le  i \le M\}$. Then $\theta_i$ has a prior gamma distribution with parameters $(\a_2,\be_2)$. Given data $\{l_i,T_i\; ; \;  1 \le  i \le M\}$ (the knowledge of $\{t_{i,j}\}$'s and $\{s_{i,j}\}$'s is not necessary here) and using Bayesian formula, we obtain that the posterior distribution of $\theta_i$ is a gamma distribution with parameters $(\a_2+l_i,\be_2+T_i)$. Thus,
\begin{align}
\what \la_i  &= \Gam(\what \a + n_i, \what \be + \tau_i),\notag\\
\what r         &= \Big(\sum_{i=1}^M n_i \Big)^{-1} \sum_{i=1}^M \tilde k_i, \label{eg10}\\
\what \th_i &= \Gam(\what \a_2 + l_i, \what \be_2 + T_i),\quad i=1,\ldots,M.\notag
\end{align}

\textbf{Model B.3.}
Parameters $(\a,\mu)$, $(\psi_1,\psi_2)$ and $(\a_2,\be_2)$ can be estimated using two-dimensional optimization procedures for functions \eqref{es3}, \eqref{es34}, \eqref{es35} respectively (see Appendix Section \ref{PARAMEST} for details).

Then, similar to \eqref{eg10}, as $\la_i$ has a prior gamma distribution with parameters $(\a,\be)$, given data $\{n_i,\tau_i\; ; \;  1 \le  i \le M\}$, the posterior distribution of $\la_i$ is a gamma distribution with parameters $(\a+n_i,\be+\tau_i)$. Correspondingly, $\theta_i$ has a prior gamma distribution with parameters $(\a_2,\be_2)$, thus, given data $\{l_i,T_i\; ; \;  1 \le  i \le M\}$, the posterior distribution of $\theta_i$ is a gamma distribution with parameters $(\a_2+l_i,\be_2+T_i)$. Furthermore,
$r_i$ has a prior beta distribution with parameters $(\psi_1,\psi_2)$, thus, given $\{n_i,\tilde k_i\; ; \;  1 \le  i \le M\}$, the posterior distribution of $r_i$ is a beta distribution with parameters $(\psi_1+\tilde k_i,\psi_2+n_i-\tilde k_i)$. To sum up, the posterior distributions of the rates of inclusion, probabilities of instantaneous dropout and rate of dropout are
\begin{align}
\what \la_i  &= \Gam(\what \a + n_i, \what \be + \tau_i), \nn \\
\what r_i     &= \Bet(\what \psi_1+\tilde k_i, \what \psi_2+n_i-\tilde k_i), \label{eg9}\\
\what \th_i &= \Gam(\what \a_2 + l_i, \what \be_2 + T_i),\quad i=1,\ldots,M.\nn
\end{align}

\section{Prediction of the number of randomized patients.} \label{prediction}

\subsection{Dropout at the time upon arrival.}
\noindent
\textbf{Model A.1.} Given data $\{(n_i,\tau_i)\; ; \;  1 \le  i \le M \}$ at interim time $t_1$, the predicted number of recruited patients $\what n_i(t)$, $t >t_1$, in center $i$ is a Poisson-gamma process with posterior rate $\what \la_i =\Gam(\what \a + n_i, \what \be + \tau_i)$ (see \eqref{eg7} and Anisimov\cite{an-comm-stat11}). Consider now predicting the number of patients that will be randomized. Denote by $(\what \mu, \what \be, \what r)$ the estimators of parameters $(\mu, \be, r)$. Recall that $R$ is a screening delay. Let $\nu_i$ be the number of patients entered screening stage at centre $i$ in the interval $[t_1-R, t_1]$ and $k_i$ be the total number of randomized patients up to time $t_1$.

Then, given $\nu_i$, the number of patients that will be randomized in the interval $[t_1, t_1+R]$ is a binomial random variable $\Bin(\nu_i,  r)$ and the times when these patients are randomized are uniformly distributed in $[t_1, t_1+R]$. The predicted number of randomized patients in  $[t_1, t_1+R]$ is $\Bin(\nu_i,  \what r)$, where $\what r$ is defined in \eqref{es32}. The number of patients randomized after time $t_1+R$ can be considered as thinning  of the process $N_\cdot^i$ with  probability  $\what r$. Let $\Pi_a$ stand for a Poisson process with rate $a$. Then, for any $t >t_1+R$, the predicted process of the number of randomized patients in centre $i$, $\{\what k_i(t), \ t\geq t_1+R \}$, is developing as a Poisson process with rate $\what r \what \la_i$. Thus,
\begin{equation} \label{pred1}
\what k_i(t) = k_i + \Bin(\nu_i, \what r) + \Pi_{\what r \, \what \la_i} (t - t_1 -R).
\end{equation}

\textbf{Model A.2.} In this case for $t >t_1+R$ we can use for predictive process $\what k_i(t)$ in centre $i$ a formula similar to \eqref{pred1}, where the posterior estimators $\what \la_i$ and $\what r_i$ are now given by \eqref{eg7}.

\subsection{Dropout during screening.} 

For models B.1-B.3, since we have more information at time $t_1$, the patients with unknown screening outcome correspond to the case $t_1-R<t_{i,j}\leq t_1$ and $Z_{ij} > t_1 - t_{i,j}$. For $1\leq i \leq M$, let $\Omega_i$ be the corresponding set of indices
\begin{equation} \label{OmegaI}
\Omega_i = \left\{  j \in \N : t_1-R<t_{i,j}\leq t_1 \text{ and } Z_{ij} > t_1 - t_{i,j}\right\}
\end{equation}
and $\nu_i = \Card(\Omega_i)$. Conditionally on $\th_i$, the probability of  such patient to be randomized in $[t_1,t_1+R]$ is  
$$
\Pcro{Z_{ij}\geq R \, | \, Z_{ij}> t_1 - t_{i,j}; \th_i} = e^{-\th_i(t_{i,j}+R-t_1)}
$$
Given data at $t_1$ and $\th_i$, the number of randomized patients between $t_1$ and $t_1+R$ in centre $i$ is the sum of $\nu_i$ independent Bernoulli r.v. with probabilities $e^{-\th_i(R-t_1+t_{i,j})}$. Denote by $\text{Ber}(p )$ a Bernoulli r.v. with probability $p$. Then the predictive process $\{\what k_i(t), t\geq t_1+R \}$ for $t>t_1+R$ can be written as
\begin{equation}  \label{pred3t}
\what k_i(t) = k_i + \Pi_{\what p_i \, \what \la_i} (t - t_1 -R) + \sum_{j \in \Omega_i} \text{Ber} \left(e^{-\th_i(t_{i,j}+R-t_1)}\right).
\end{equation}
The probability of non-dropout is $\what p_i = \what r_i\exp(-\what \th_i R)$. \\

\textbf{Model B.1.} In this case $\what r_i \equiv \what r$ and $\what \th_i \equiv \what \th$,  where $(\what r,\what \th)$ are given in \eqref{es33}. Thus,  using \eqref{pred3t} we get
$$
\what k_i(t) = k_i + \Pi_{\what p \, \what \la_i} (t - t_1 -R)  + \sum_{j \in \Omega_i} \text{Ber} \left(e^{-\what \th(t_{i,j}+R-t_1)}\right),
$$
where $\what p = \what r e^{-\what \th R}$. \\

\textbf{Model B.2.} The predictive process in centre $i$ is
$$
\what k_i(t) = k_i + \Pi_{\what p_i \, \what \la_i} (t - t_1 -R)
 + \sum_{j \in \Omega_i} \text{Ber} \left(e^{-\what \th_i(t_{i,j}+R-t_1)} \right) ,
$$
where $\what p_i = \what r \exp(-\what \th_i R)$, and $\what \la_i, \what r$ and $\what \th_i$ are given by \eqref{eg10}. \\

\textbf{Model B.3.} In this case $\what r_i$ is also random and given in \eqref{eg9}.\\

Denote by $\what k(t) = \sum_{i=1}^M \what k_i(t)$ the total number of randomized patients at time $t > t_1+R$. For each model, at large enough $M$ ($M>10$) we can use the expressions  of the expectation and the variance of $\what k(t)$ given the data  to create $(1-\de)$-predictive bounds for $\what k(t)$ using a normal approximation similar to the method used in Anisimov \cite{an-comm-stat11}. These expressions are given for each model in Appendix Section \ref{S-APPREC}.

\begin{remark}
 It is also possible to consider a joint distribution of the two-component process $\{ (\what n_i(t),\what k_i(t)), \ t \geq t_1+R \}.$ Given data at time $t_1$, in interval $[t_1,t_1+R]$ these processes are independent, as for $t \in [t_1,t_1+R]$ the process $\what k_i$ depends only on the data before time $t_1$. For $t > t_1+R $, the process $\{\what k_i(t),\ t\geq t_1+R \}$ can be represented as thinning of the process $\{\what n_i(t-R),\ t\geq t_1\}$ with probability $\what r$.
\end{remark}

\section{Simulation studies.} \label{simulation}

Simulation studies are split into two parts. A first part is devoted to the investigation of the models for dropout upon arrival (models A.1-A.2) and a second part to investigate the models for dropout during a fixed screening period $R$ (models B.1-B.3).

\subsection{Data generation procedure.} \label{S-DG}

\subsubsection{Part I: models for dropout upon arrival.}

We simulate data according to model A.2. One has to generate the recruitment rates $\la_i$ of the centres according to a Ga$(\a,\be)$ distribution and the probabilities of staying in trial at arrival $r_i$ according to a Beta$(\psi_1,\psi_2)$ distribution. Then, the inter-arrival times between patients entering at centre $i$ are exponential variables with parameter $\la_i$. With each inclusion time $t_{i,j}$ in centre $i$, it is associated a Ber($r_i$) random variable denoted $\chi_{ij}$. In the following, $\mu = \a / \be$ denotes the mean rate per centre.

In order to be close to what is observed in practice \cite{an-fed07, MS12, Minois1, Minois2}, we have chosen the expected number of randomized patients at the end of the recruitment period $N=750$, the number of centres $M=75$ and a mean recruitment rate of $\mu = 3.5$ patients per year and per center. Finally we have choosen an average instantaneous dropout rate of 0.2. The theoretical trial duration is thus 3.57 years. A set of parameters coherent with this data is given in Table \ref{tablecoeff}.\\

\subsubsection{Part II: models for dropout during screening.}

We simulate data according to model B.3. One has to generate the rates $\la_i$ of the centres according to a Ga$(\a,\be)$ distribution, the probabilities of staying in trial at arrival $r_i$ according to a Beta$(\psi_1,\psi_2)$ distribution, and the rates $\theta_i$ of the exponential durations according to a Ga$(\a_2,\be_2)$ distribution. Then, the inter-arrival times between patients entering at centre $i$ are exponential variables with parameter $\la_i$. With each inclusion time $t_{i,j}$ in centre $i$, it is associated an exponential time $Z_{ij}$ with rate $\theta_i$ and a Ber($r_i$) random variable denoted $\chi_{ij}$.

As in the previous setting we have choosen the expected number of randomized patients at the end of the recruitment period $N=750$, the number of centres $M=75$, a mean recruitment rate of $\mu = 3.5$ patients per year per center and an average instantaneous dropout rate of 0.2. Finally we have chosen a screening duration $R=0.2$ year and a mean dropout during screening rate of 0.33. The theoretical trial duration is thus 5.33 years. A set of parameters coherent with this data is given in Table \ref{tablecoeff}.

\begin{remark}
Notice that we have chosen to consider the same values of $\alpha$ and $\beta$ for both parts, the duration of the trial is thus longer for part II compared with part I due to possible dropout during screening period.
\end{remark}

\begin{table}
\caption{Sets of parameters coherent with the chosen data and used for data generation.} \label{tablecoeff}
\centering
\begin{tabular}{l c c}
\hline
 										&Part I 		&Part II  \\
\hline
Number of centres						&$M = 75$		&$M = 75$\\
Number of patients to be recruited		&$N = 750$		&$N = 750$\\
Recruitment								&$\a = 1.2$		&$\a = 1.2$\\
										&$\mu = 3.5$	&$\mu = 3.5$\\
Instantaneous dropout 					&$\psi_1 = 4$ 	&$\psi_1 = 4$\\
										&$\psi_2 = 1$	&$\psi_2 = 1$\\
Screening duration						&NA				&$R = 0.2$\\
Dropout during screening					&NA				&$\a_2 = 1$\\
										&NA				&$\mu_2 = 2$ \\
\hline
\end{tabular}
\end{table}

\subsection{Simulation scenario.} \label{S-SS}
We generate data as specified in section \ref{S-DG} for Parts I and II. For each part, using simulated data, we investigate two main questions of interest:
\begin{itemize}
\item Choosing the interim times $t_1=t_1^1, t_1^2,  t_1^3$, one collects for each center $i$ the required data for each model and estimates the different values of the parameters. That allows us to evaluate the behavior of the parameters as $t_1$ is changing.
\item Choosing the interim times $t_1=t_1^1, t_1^2,  t_1^3$, we estimate the duration of the trial in order to evaluate the sensitivity of this estimation with respect to $t_1$.
\end{itemize}
We choose the interim times $t_1 = 1, 1.5, 2$ years for Part I (models A.1 and A.2) and  $t_1 = 1$, $2, 3$ years for Part II (models B.1-B.3).
For each part and interim time, the results described in the following sections are based on 5000 data replications.

\subsection{Parameters estimation.}

\begin{figure}[p]
 \centering
\includegraphics[width=15cm]{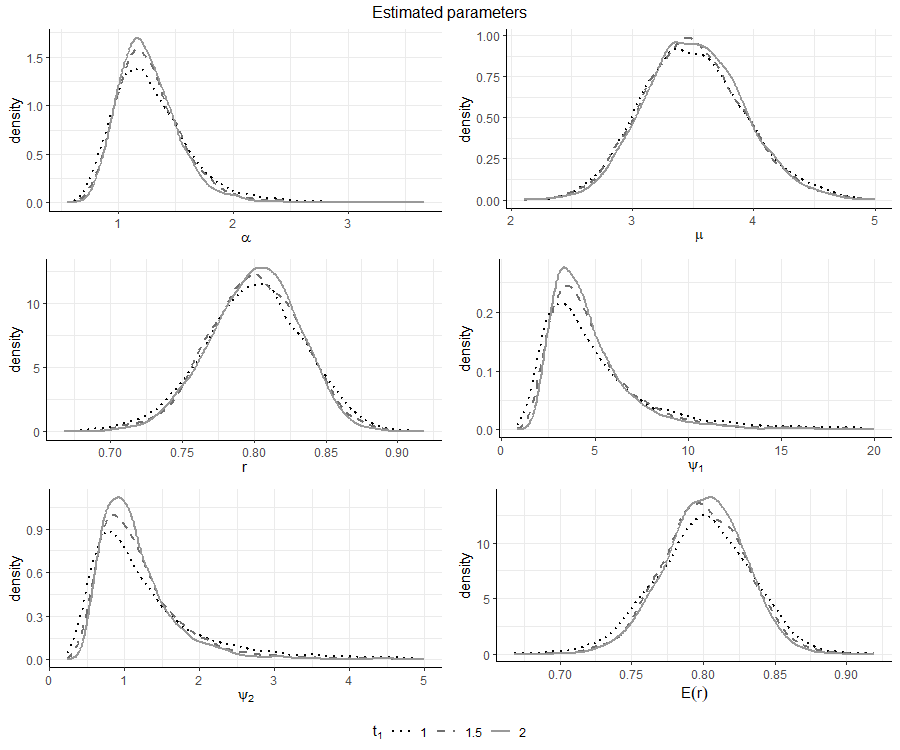}
\caption{Empirical distribution of the estimated parameters involved in models A.1-A.2 based on 5000 data replications for different values of the interim time $t_1$.}
\label{par_est_1}
\end{figure}

\begin{figure}[p]
 \centering
\includegraphics[width=15cm]{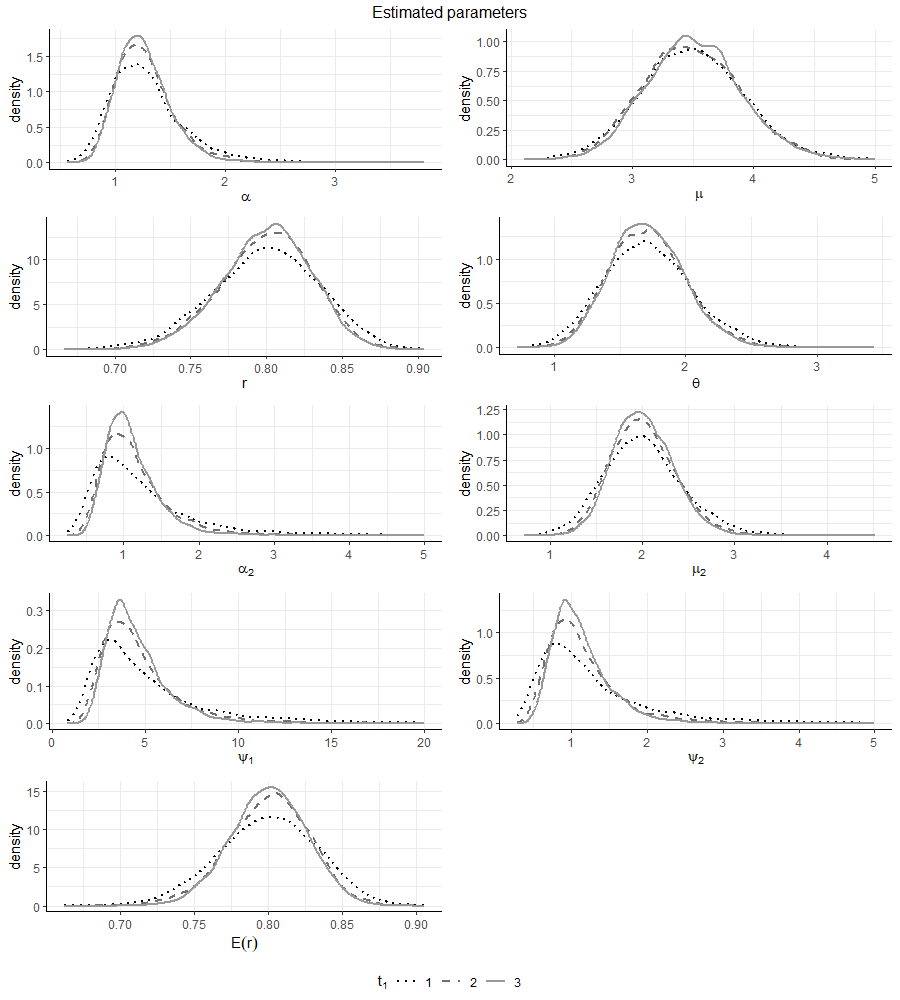}
\caption{Empirical distribution of the estimated parameters involved in models B.1-B.3 based on 5000 data replications for different values of the interim time $t_1$.}
\label{par_est_2}
\end{figure}

For the recruitment process, the estimated parameters at different interim times $t_1$ are given in Table \ref{table1}. Furthermore, a graphical representation of these estimates across data replications is depicted in Figure \ref{par_est_1} for models A.1-A.2 and Figure \ref{par_est_2} for models B.1-B.3. Note that $\a$ and $\mu$ are estimated for all models, but for the illustrative purpose of Table \ref{table1} the reported estimates are the ones related to models A.1 and A.2. No relevant difference was observed in the estimation of both parameters under models B.1-B.3.\\

\begin{table}
\caption{Monte Carlo average and standard deviation (in brackets) of the estimated model parameters across 5000 data replications as a function of $t_1$ ($t_1 = 1, 1.5, 2$ years for Part I (models A.1 and A.2) and  $t_1 = 1$, $2, 3$ years for Part II (models B.1-B.3)).} \label{table1}
\centering
\begin{tabular}{c c c c c c c}
\hline
 &\multirow{2}{*}{Parameters} &Models &\multicolumn{3}{c}{Interim time $t_1$}\\
  &                     &Involved & $t_1^1$ & $t_1^2$ & $t_1^3$ \\
\hline
\multirow{2}{2cm}{Recruitment} &$\what \a$ & All & 1.29 (0.34) & 1.28 (0.29) & 1.26 (0.26) \\
&$\what \mu$ & All & 3.50 (0.43) & 3.50 (0.41) & 3.51 (0.40) \\
\hline
\multirow{4}{2cm}{Dropout at inclusion}
&$\what r$ & A.1 & 0.80 (0.04) & 0.80 (0.03) & 0.80 (0.03)\\
&$\what \psi_1$ & A.2 & 7.70 (18.92) & 5.36 (7.20) & 4.83 (2.96) \\
&$\what \psi_2$ &A.2 & 1.81 (4.19) & 1.30 (1.68) & 1.17 (0.64) \\
&$\what \psi_1/(\what \psi_1+\what \psi_2)$ & A.2 & 0.80 (0.03) & 0.80 (0.03) & 0.80 (0.03)\\
\hline
\multirow{7}{2cm}{Dropout during screening period}
&$\what \th$ &B.1 & 1.72 (0.34) & 1.71 (0.29) & 1.71 (0.28)\\
&$\what r$ & B.1-B.2 & 0.80 (0.04) & 0.80 (0.03)& 0.80 (0.03)\\
&$\what \a_2$  &B.2-B.3 & 26.62 (422.09) & 1.15 (0.50) & 1.1 (0.37)\\
&$\what \mu_2$ &B.2-B.3 & 2.02 (0.43) & 2.01 (0.35) & 2.01 (0.32)\\
&$\what \psi_1$ &B.3 & 27.00 (277.16) & 5.90 (70.94) & 4.58 (1.80)\\
&$\what \psi_2$ &B.3 & 5.90 (59.47) & 1.42 (15.55)& 1.12 (0.40)\\
&$\what \psi_1/(\what \psi_1+\what \psi_2)$ &B.3 & 0.80 (0.03) & 0.80 (0.03)&  0.80 (0.03)\\
\hline
\end{tabular}
\end{table}

First, we can see that the estimated quantities approach their target values as $t_1$ increases. The estimates of some parameters, namely $\what \psi_1, \what \psi_2 $ (in both  models A.2 and B.3) and $\what \a_2$,  show a high level of variability for the first interim time and their average seems off target. However, $\widehat{\Esp[r_i]}=\frac{\what \psi_1}{\what \psi_1+\what \psi_2}$ (in both models  A.2 and B.3) and $\widehat{\Esp[\th_i]} =\what \mu_2$ remain stable across values of $t_1$ and their average across data replications is remarkably close to their target values. The estimates of the other parameters are fairly stable and close to their original values used to generate the data.

\subsection{Trial duration estimation.}
\begin{figure}
 \centering
\includegraphics[width=15cm]{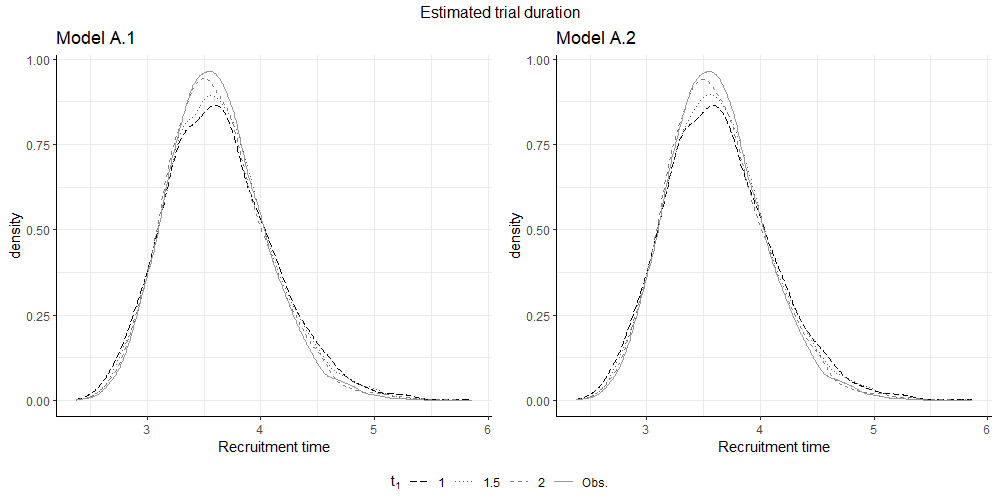}
\caption{Empirical distributions of the estimated recruitment time under models A based on 5000 data replications for different values of the interim time $t_1$ compared with the observed duration.}
\label{fig_time_A}
\end{figure}

\begin{figure}
 \centering
\includegraphics[width=15cm]{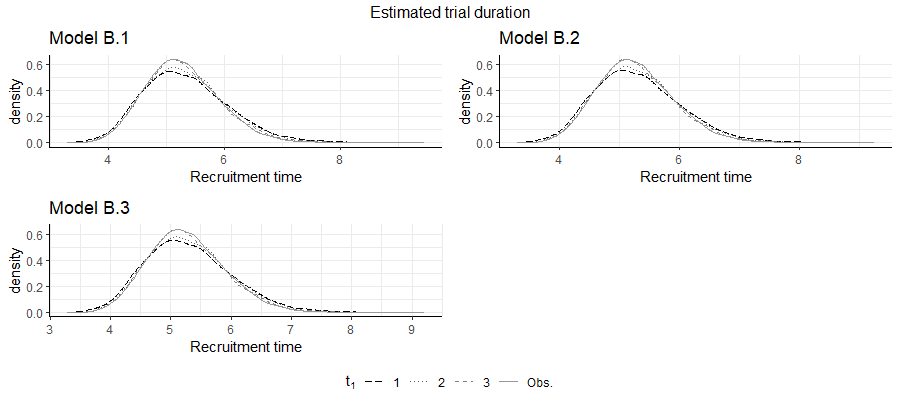}
\caption{Empirical distributions of the estimated recruitment time under models B based on 5000 data replications for different values of the interim time $t_1$ compared with the observed duration.}
\label{fig_time_B}
\end{figure}

In Tables \ref{tableesp1} and \ref{tableesp2} we show the estimated recruitment times under models A.1-A.2 and B.1-B.3 respectively for different interim times. In the setting where the dropout process takes place only upon arrival, i.e. models A.1 and A.2, the point estimates of the recruitment time under the two models are nearly identical. However, model A.2 performs better in terms of coverage rate of the observed recruitment time by the $95\%$ credible interval. This is expected, as model A.2 allows for the probability of randomization $r_i$ to vary across centers, allowing for a more realistic estimation of the variability around the point estimate. On the other hand, in the setting which entails a screening process, that are models  B.1-B.3, the differences between models are slightly more evident. Specifically, models B.2 and B.3 both outperform model B.1, especially for the first interim time $t_1=1$. This is not surprising, as model B.1 does not account for the variation across centers in the probability of dropout upon arrival and the distribution of dropout times during the screening process. Models B.2 and B.3 perform similarly, with model B.3 leading to marginally better results. A visual representation that illustrates the accuracy of models is shown in Figure \ref{fig_time_A} for models A and Figure \ref{fig_time_B} for models B. Since the major difference between models boils down to how they account for the uncertainty around point estimates, the same plots but for the remaining models are almost identical.

\begin{table}
\caption{Monte Carlo average, standard deviation, percent bias, and coverage rate of the $95\%$ CI of the observed recruitment time across 5000 data replications for different interim times $t_1$ for models A. The average observed recruitment times was 3.62 (SD=0.42) years.}
\label{tableesp1}
\centering
\begin{tabular}{c c c c c c }
\hline
Model  & $t_1$& Mean & SD & \% Bias & Cov. \\
\hline                  
\multirow{3}{*}{A.1}
 &1    &3.64   &0.48  &4.63 &0.90\\ 
 &1.5  &3.64   &0.46  &3.50 &0.91\\
 &2    &3.61   &0.44  &2.61 &0.93\\
 \hline
\multirow{3}{*}{A.2}
 &1    &3.64   &0.48  &4.63 &0.92\\ 
 &1.5  &3.63   &0.45  &3.50 &0.93\\
 &2    &3.61   &0.44  &2.61 &0.94\\
\hline
\end{tabular}
\end{table}

\begin{table}
\caption{Monte Carlo average, standard deviation, percent bias, and coverage rate of the $95\%$ CI of the observed recruitment time across 5000 data replications for different interim times $t_1$ for models B. The average observed recruitment times was 5.27 (SD=0.64) years.}
\label{tableesp2}
\centering
\begin{tabular}{c c c c c c }
\hline
Model  & $t_1$& Mean & SD & \% Bias & Cov. \\
\hline                  
\multirow{3}{*}{B.1}
&1  & 5.34 &0.76&5.79&0.84 \\ 
&2& 5.33 &0.71&3.58& 0.89\\
&3  & 5.29& 0.67&2.36& 0.91 \\
 \hline
\multirow{3}{*}{B.2}
&1 &5.31& 0.75&5.65& 0.88\\ 
&2&5.31&0.70&3.51& 0.93\\
&3  &5.28 &0.66&2.32& 0.94\\
 \hline
\multirow{3}{*}{B.3}
&1 &5.31& 0.74&5.65& 0.89 \\ 
&2&5.31& 0.70&3.49& 0.93 \\
&3&5.28&0.66&2.31&  0.94\\
\hline
\end{tabular}
\end{table}

\section{Conclusion.} \label{conclu}

In this paper, a new methodology for interim re-projecting patient recruitment in multicenter clinical trials accounting for various types of dropout is developed. Specifically, five different models belonging to two separate categories have been presented: the models that belong to the first group (A.1 and A.2) only consider patient dropout upon arrival, whereas the ones in the second group (B.1, B.2, and B.3) also take into account dropout during the screening period which follows the patients’ arrival at the centers. Within each category, the difference between models lies in the level of variability of model parameters across centers that is considered. The technique for estimating parameters using interim data and predicting the number of recruited/randomized patients over time and the recruitment time is developed. This methodology is validated using a simulation study. The results show that all five models reach a satisfactory performance as the interim time and the proportion of already randomized patients increase. However, the models with higher complexity in their respective categories, i.e. models A.2 and B.2-B.3, achieve better results in terms of the credible interval coverage probabilities of the observed trial duration as they allow for a more realistic estimation of the variability in the probability of screening failures across different centers. Nevertheless, despite the vast applicability of this methodology to real clinical trials data, the necessary information on dropout is rarely collected in practice. Therefore, one of the important aims of this paper is to emphasize the need for a more thorough collection of this type of data to allow for more precise and realistic estimates and re-projection of the recruitment process in clinical trials.

\section*{Acknowledgments}
Authors thank Sandrine Andrieu, Nathan Minois and St\'ephanie Savy for valuable discussions on this topic. This research has received the help from IRESP during the call for proposals launched in 2012 as a part of French "Cancer Plan 2009-2013".

\subsection*{Conflict of interest}
The authors declare no potential conflict of interests.

\renewcommand{\thesection}{\Alph{section}}
\setcounter{section}{0}

\section{Appendix}

\subsection{Parameters' estimation.} \label{PARAMEST}
Let us recall some notations. Models A (A.1 and A.2) and the parameters involved ($r$, $(\Psi_1,\Psi_2)$) are defined in Section \ref{S-DOUA}. Models B (B.1, B.2 and B.3) and the parameters involved ($\theta$, $r$, $(\alpha_2,\mu_2)$, $(\Psi_1,\Psi_2)$) are defined in Section \ref{S-DODS}.
Finally, $(\alpha,\mu)$ are the parameters of the Poisson-gamma recruitment model defined in Section \ref{estimation}.

To estimate these parameters at a given interim time $t_1$, consider, for some centre $i$, $\tau_i=t_1-u_i$ (where $u_i$ stands for centre $i$ opening date), $n_i$ the number of patients recruited and $k_i$ the number of patients randomized at time $t_1 > u_i$. If there is a screening period $R$ (Models B), then assume that $\tau_i \geq R$ and the observed data must be enriched by considering $l_i$ the number of patients lost during the screening process, $T_i= \sum_j (s_{i,j}-t_{i,j})$  the sum of screening durations, and $\tilde k_i$ the number of patients that are not lost immediately upon arrival and for Model B.3 by considering the patients arrival times $\{t_{i,j}\leq t_1\; ;\; j \ge 1\; , \;  1 \le  i \le M \}$ and the last times $\{s_{i,j}, j \ge 1\; ; \;  1 \le  i \le M	\}$  they were in the screening process (i.e $t_{i,j} \le s_{i,j} \le t_1$).\\

\noindent
\textbf{Model A.1.} Given data $\{n_i, k_i, \tau_i\}$ at the interim time $t_1$, the log-likelihood function can be written in the form:
$$
\cL_1(\a, \mu, r) = \sum_{i=1}^M \ln \left[{\rm NegBin}\left(n_i;\a, \frac{\mu\tau_i}{\a+\mu\tau_i}\right)\right] + \sum_{i=1}^M  \ln[{\rm Bin}(k_i;n_i,r)].
$$
As we see, the parameter $r$ is separated from $(\a,\mu)$ and $\cL_1(\a, \mu, r)$ can be re-written in the form: $\cL_1(\a, \mu, r)= \cL_{1,1}(\a, \mu)+\cL_{1,2}(r),$
with
\begin{equation} \label{es3}
\cL_{1,1}(\a, \mu) = \sum_{i=1}^M \ln \Ga(n_i+\a) - M \ln \Ga(\a) + N_1 (\ln \mu - \ln \a) - \sum_{i=1}^M (n_i+\a) \ln(1+\mu \tau_i/\a) +C,
\end{equation}
and
\begin{equation*}
\cL_{1,2}(r) = \sum_{i=1}^M  \Big[ k_i \ln r + (n_i - k_i) \ln (1-r)  \Big] +C,
\end{equation*}
where $C$ is some generic constant independent of the parameters, and $N_1=\sum_{i=1}^M n_i$ is the total number of  recruited patients up to time $t_1$. Taking derivative in $r$ it is easy to calculate that the maximum likelihood estimator is given by \eqref{es32}.\\

\textbf{Model A.2.} Given data $\{n_i, k_i, \tau_i\}$ at an interim time $t_1$, the log-likelihood function can be written in the form:
$$
\cL_2(\a, \mu, \psi_1,\psi_2)= \cL_{2,1}(\a, \mu)+\cL_{2,2}(\psi_1,\psi_2),
$$
where $\cL_{2,1} =  \cL_{1,1}$ given by \eqref{es3}, and
\begin{equation} \label{eq6}
\cL_{2,2}(\psi_1,\psi_2)=
\sum_{i=1}^M \ln \cB(k_i+\psi_1, n_i-k_i+\psi_2) - M \ln \cB(\psi_1,\psi_2) +C.
\end{equation}

\textbf{Model B.1-B.3.}  Conditioning on parameters $\{ \theta_i, r_i \, ; \, 1\leq i \leq M\}$, we can write a general expression for the likelihood
\begin{align}
\mathbf L\left[ (t_{i,j}); (s_{i,j}) \right] = \exp\left[ \cL_{1,1} (\a,\mu) \right]\times \prod_{i=1}^M \Esp\left[ \mathbf L_2\left(\th_i, r_i ; (t_{i,j}),(s_{i,j})\right) \right]\label{egl}
\end{align}
where $\cL_{1,1}$ is given in \eqref{es3}, and
\begin{align*}
\mathbf L_2\left(\th_i, r_i ; (t_{i,j}),(s_{i,j})\right)
&=  \prod_{\cD_1}  (1-r_i)  ~ \prod_{\cD_2} r_i \exp(-\th_i(s_{i,j}-t_{i,j}) ) \\
& \qquad \times\prod_{\cD_3} r_i\theta_i\exp\left( - \th_i(s_{i,j}-t_{i,j}) \right) ,
\end{align*}
where $\cD_1=\bigcup_{i=1}^M \cD_1^i$, $\cD_2=\bigcup_{i=1}^M \cD_2^i$, $\cD_3=\bigcup_{i=1}^M \cD_3^i$ and for any $1 \le  i \le M$,
\begin{align*}
\cD_1^i &= \{ j \ge 1, \, \text{such that} \quad s_{i,j}=t_{i,j} \},\\
\cD_2^i &= \{ j \ge 1, \, \text{such that} \quad s_{i,j}=(t_{i,j}+R)\wedge t_1 \},\\
\cD_3^i &= \{ j \ge 1, \, \text{such that} \quad t_{i,j}<s_{i,j}<(t_{i,j}+R)\wedge t_1 \}.
\end{align*}

In \eqref{egl}, the expectation is taken when $\th_i$ and $r_i$ vary according to their respective distributions defined by models B.1-B.3.

Notice that $l_i = \Card \,( \cD_3^i)$ and $\tilde k_i = \Card \,( \cD_2^i )+\Card \,( \cD_3^i )$. Then $\mathbf L_2(\cdot)$ can be rewritten as
\begin{equation} \label{egl2}
\mathbf L_2\left(\th_i, r_i ; (t_{i,j}),(s_{i,j})\right) = (1-r_i)^{n_i-\tilde k_i} r_i^{\tilde k_i} \theta_i^{l_i}\exp\left( -\theta_i T_i \right).
\end{equation}

\textbf{Model B.1.} Given data $\{t_{i,j},s_{i,j}, \tau_i, j \ge 1\; ; \;  1 \le  i \le M\}$ at an interim time $t_1$, and using \eqref{egl} and \eqref{egl2}, the log-likelihood function can be written in the form:
$$
\cL_3(\a,\mu,r,\th)= \cL_{3,1}(\a, \mu)+\sum_{i=1}^M  \Big[(n_i - \tilde k_i) \ln (1-r) + \tilde k_i\ln r  + l_i \ln \th - \th T_i\Big] ,
$$
where $\cL_{3,1} = \cL_{1,1}$ is given in \eqref{es3} and $(\what \a, \what \mu)$ can be calculated using a two-dimensional optimization procedure for function $\cL_{1,1}(\a,\mu)$. Taking derivatives in $r$ and $\th$ we get the maximum likelihood estimators \eqref{es33}.\\

\textbf{Model B.2.} Given data $\{t_{i,j},s_{i,j}, \tau_i, j \ge 1\; ; \;  1 \le  i \le M\}$ at an interim time $t_1$, and using \eqref{egl} and \eqref{egl2}, the log-likelihood function can be written in the form:
$$
\cL_4(\a,\mu,r,\a_2, \be_2) =\,\cL_{4,1}(\a, \mu) + \sum_{i=1}^M  \Big[ (n_i - \tilde k_i) \ln (1-r) + \tilde k_i \ln r \Big] +\cL_{4,2}(\a_2,\be_2)
$$
where
\begin{equation} \label{es34}
\cL_{4,2} (\a_2,\be_2)= \sum_{i=1}^M \bigg[ \ln \Ga(l_i+\a_2)-  \ln \Ga(\a_2) + \a_2 \ln \be_2- (l_i+\a_2) \ln(\be_2+ T_i)\bigg]
\end{equation}
and $\cL_{4,1} = \cL_{1,1}$ is given by \eqref{es3}. $(\what \a, \what \mu)$ and $(\what \a_2, \what \beta_2)$ are calculated numerically using two-dimensional optimization procedure, and, by derivation, $\what r$ is given by \eqref{es33}.\\

\textbf{Model B.3.} Given data $\{t_{i,j},s_{i,j}, \tau_i, j \ge 1\; ; \;  1 \le  i \le M\}$ at an interim time $t_1$, and using \eqref{egl} and \eqref{egl2}, the log-likelihood function can be written in the form:
$$
\cL_5(\a,\mu,\psi_1,\psi_2, \a_2,\be_2)  =\cL_{5,1}(\a, \mu) +\cL_{5,2}(\a_2,\be_2) + \cL_{5,3}(\psi_1,\psi_2)
$$
where
\begin{equation} \label{es35}
\cL_{5,3}(\psi_1,\psi_2)= \sum_{i=1}^M \bigg[ \ln \cB(\tilde k_i+\psi_1,n_i-\tilde k_i+\psi_2) -  \ln \cB(\psi_1,\psi_2) \bigg] ,
\end{equation}
$ \cL_{5,1} =  \cL_{1,1}$ is given by \eqref{es3} and $\cL_{5,2} = \cL_{4,2}$ is given by \eqref{es34}. Parameters $(\a,\mu)$, $(\psi_1,\psi_2)$ and $(\a_2,\be_2)$ can be estimated using two-dimensional optimization procedures for corresponding functions $\cL(\cdot)$.

\subsection{Prediction of the number of randomized patients.}\label{S-APPREC}
~~\\
~~\\
Let $K_2 = \sum_{i=1}^M k_i$.\\

\textbf{Model A.1.}  Note that for a random rate $\la$, $\Esp [\Pi_{\la}(t)]=t \Esp[\la]$, $\Var [\Pi_{\la}(t)]=t \Esp[\la]+t^2 \Esp[\la]$. Therefore, given interim data,
$$
\Esp [\what k_i(t) \mid \data] = k_i + \nu_i \what r + \what r \, (t - t_1 -R)\Esp[\what \la_i] ,
$$
where $\what r$ is given by \eqref{es32}, and
\begin{equation} \label{mvarla}
\Esp[\what \la_i] = \frac {\a+n_i} {\be + \tau_i}, \qquad \Var[\what \la_i] = \frac {\a+n_i} {(\be + \tau_i)^2}.
\end{equation}
As for given $\{n_i, k_i \}$ the posterior predictors $\what \la_i$ are independent of $\what r$, then the variance is
\begin{equation}  \label{eq4}
\Var [\what k_i(t) \mid \data] = \nu_i \what r(1-\what r) + \what r (t - t_1 -R)\Esp[\what \la_i] + \what r^2 (t - t_1 -R)^2 \Var[ \what \la_i ].
\end{equation}
Finally,
\begin{align*}
\Esp [\what k(t) \mid \data] &= K_2 + \what r \, \sum_{i=1}^M \nu_i + \what r \, (t - t_1 -R) \sum_{i=1}^M \frac {\a+n_i} {\be + \tau_i},\\
\Var [\what k(t) \mid \data] &= \what r(1-\what r) \, \sum_{i=1}^M \nu_i + \what r \, (t - t_1 -R) \sum_{i=1}^M \frac {\a+n_i} {\be + \tau_i} \\
														& \qquad  + \what r^2 (t - t_1 -R)^2 \sum_{i=1}^M \frac {\a+n_i} {(\be + \tau_i)^2}.
\end{align*}

\textbf{Model A.2.}  The mean and the variance of $\la_i$ are calculated in \eqref{mvarla}, and
$$
\Esp[\what r_i \mid \data] = \frac {\psi_1+k_i} {\psi_1 + \psi_2 +n_i}, \qquad \Var [\what r_i \mid \data] =
\frac {(\psi_1+k_i)(\psi_2+n_i-k_i)} {(\psi_1 + \psi_2+n_i)^2(1+\psi_1 + \psi_2 +n_i)},
$$
where instead of $(\psi_1, \psi_2)$ we should substitute $(\what \psi_1, \what \psi_2)$. As data $\{n_i, k_i \,;\, 1\leq i\leq M\}$ are given, the posterior predictors $\what \la_i$ and $\what r_i$ are independent. Note that for a random probability $r$, $\Esp [\Bin(n, r)] = n \Esp [r]$, and
\begin{equation*}
\Var [\Bin(n, r)] = \Esp[ \Var [\Bin(n, r) \mid r]  ]+ \Var[ \Esp [\Bin(n, r) \mid r]  ] = n \Esp [r(1-r)]+ n^2 \Var [r].
\end{equation*}
Therefore,
\begin{equation*}
\Esp [\what k_i(t) \mid \data] = k_i + \nu_i \Esp[\what r_i] + (t - t_1 -R)\Esp[\what r_i] \Esp[\what \la_i] ,
\end{equation*}
and
\begin{equation*} \label{eq9}
\Var [\what k_i(t) \mid \data] = \nu_i \Esp [\what r_i(1-\what r_i)] + \nu_i^2 \Var [\what r_i]  \nn  +
(t - t_1 -R) \Esp[\what r_i] \Esp[\what \la_i] + (t - t_1 -R)^2 \Var[\what r_i \what \la_i ],
\end{equation*}
where we can use the formula
$$
\Var[\what r_i \what \la_i ]= \Esp[ \what \la_i^2] \Var [\what r_i] +
(\Esp[ \what r_i])^2 \Var [\what \la_i].
$$
Finally, using relations \eqref{mvarla}, \eqref{eq4} we can easy calculate the mean and the
variance of the global process $\what k(t)$.\\

\textbf{Model B.1.}  We have
\begin{align*}
\Esp[ \what k_i(t) \mid \data ] &= k_i + (t-t_1-R)  \what r e^{-\what \th R} \Esp [\what \la_i]+ \sum_{j \in \Omega_i} g_{i,j}, \\
\Var [ \what k_i(t)  \mid \data] &=  (t-t_1-R)^2  \what r^2 e^{-2 \what \th R}
\Var[\what \la_i]+ (t-t_1-R) \what r e^{-\what \th R} \Esp[\what \la_i] \\
&\qquad  + \sum_{j \in \Omega_i} g_{i,j}~(1-g_{i,j}),
\end{align*}
where $g_{i,j} = e^{-\what \th(t_{i,j}+R-t_1)}$, $\hat{\theta}$ is given by \eqref{es33}  and $\Esp[\what \la_i]$ and $\Var[\what \la_i]$ are given in \eqref{mvarla}.\\

\textbf{Model B.2.}  Denote by $F_i(s)$ a Laplace transformation of $\what \th_i $:
\begin{equation*} \label{lapth}
F_i(s)=\Esp[e^{-\what \th_i s} \mid \data] = [1+s/(\what \be_2 + T_i)]^{-\what \a_2-l_i}, \qquad s\geq 0,
\end{equation*}
Then,
\begin{align*}
\Esp[ \what k_i(t)  \mid \data] &= k_i + (t-t_1-R)  \what r F_i(R)\Esp [\what \la_i]\nn +  \sum_{j \in \Omega_i} F_i(t_{i,j}+R-t_1),\\
\Var [ \what k_i(t) \mid \data ]& =  (t-t_1-R)^2  \what r^2  \Var[e^{- \what \th_i R}\what \la_i] + (t-t_1-R) \what r \Esp[e^{-\what \th_i R}] \Esp[\what \la_i] \\
&\qquad + \Var\left[ \sum_{j\in\Omega_i}\text{Ber}  \left( g_{i,j}  \right)  \right],
\end{align*}
where $\Esp[\what \la_i]$ and $\Var[\what \la_i]$ are given in \eqref{mvarla}, $\Var[e^{- \what \th_i R}\what \la_i] = F_i(2R)\Esp[\what\la_i^2] - (F_i(R) \Esp[\what \la_i])^2$, $\Esp[e^{-\what \th_i R}] =F_i(R)$, and straightforward calculations show that, by denoting $\Delta_{ij} = R-t_1-t_{i,j}$,
\begin{align*}
\Var\left[ \sum_{j\in\Omega_i}\text{Ber}  \left( g_{i,j}  \right)  \right]  &=  \sum_{j\in\Omega_i} F_i(\Delta_{ij}) - F_i(2\Delta_{ij}) \\
&\qquad + \sum_{(j_1,j_2) \in \Omega_i^2} F_i(\Delta_{ij_1}+\Delta_{ij_2}) - F_i(\Delta_{ij_1})F_i(\Delta_{ij_2}).
\end{align*}

\textbf{Model B.3.} In a similar way as in model B.2, we can write
\begin{align*}
\Esp[ \what k_i(t)  \mid \data] &= k_i + (t-t_1-R)  \Esp[\what r_i] F_i(R)\Esp [\what \la_i]\nn
+ \sum_{j \in \Omega_i} F_i(t_{i,j}+R-t_1),\\
\Var [ \what k_i(t)  \mid \data]& =  (t-t_1-R)^2  \Var[\what r_i e^{- \what \th_i R}\what \la_i] + (t-t_1-R) \Esp[\what r_i]\Esp[e^{-\what \th_i R}] \Esp[\what \la_i] \\
& \qquad + \Var\left[ \sum_{j\in\Omega_i}\text{Ber}  \left( g_{i,j}  \right)  \right],
\end{align*}
and use previous formulae for calculation the expectations and variances of different variables in this expression.



\begin{thebibliography}{10}

\bibitem{senn97}
Stephen Senn.
\newblock {\em Statistical Issues in Drug Development}.
\newblock John Wiley \& Sons, Chichester, 1997.

\bibitem{senn98}
Stephen Senn.
\newblock Some controversies in planning and analysing multi-centre trials.
\newblock {\em Statistics in Medicine}, 17:1753--1765, 1998.

\bibitem{cart-BMC05}
Rickey~E. Carter, Susan~C. Sonne, and Kathleen~T. Brady.
\newblock Practical considerations for estimating clinical trial accrual
  periods: application to a multi-center effectiveness study.
\newblock {\em BMC Medical Research Methodology}, 5(11):1--5, 2005.

\bibitem{cart-CCT04}
Rickey~Edward Carter.
\newblock Application of stochastic processes to participant recruitment in
  clinical trials.
\newblock {\em Controlled Clinical Trials}, 25(5):429--436, 2004.

\bibitem{bates}
Grace~E. Bates and Jerzy Neyman.
\newblock Contributions to the theory of accident proneness. {II}. {T}rue or
  false contagion.
\newblock {\em Univ. California Publ. Statist.}, 1:255--275, 1952.

\bibitem{an-fed-chapter-07}
Vladimir~V. Anisimov and Valerii~V. Fedorov.
\newblock {\em Design of multicentre clinical trials with random enrolment},
  chapter~25, pages 387--400.
\newblock Advances in Statistical Methods for the Health Sciences. Birkhauser,
  2007.

\bibitem{an-fed07}
Vladimir~V. Anisimov and Valerii~V. Fedorov.
\newblock Modelling, prediction and adaptive adjustment of recruitment in
  multicentre trials.
\newblock {\em Statistics in Medicine}, 26(27):4958--4975, 2007.

\bibitem{an-outs09}
Vladimir~V. Anisimov.
\newblock Recruitment modeling and predicting in clinical trials.
\newblock {\em Pharmaceutical Outsourcing}, 10(1):44--48, 2009.

\bibitem{ADF07}
Vladimir~V. Anisimov, D.~Downing, and Valerii~V. Fedorov.
\newblock Recruitment in multicentre trials: prediction and adjustment.
\newblock In {\em mODa 8 - Advances in Model-Oriented Design and Analysis},
  pages 1--8. Physica-Verlag HD, 2007.

\bibitem{an-JSM09}
Vladimir~V. Anisimov.
\newblock Predictive modelling of recruitment and drug supply in multicenter
  clinical trials.
\newblock In {\em Proceedings of the Joint Statistical Meeting, ASA}, pages
  1248--1259, Washington, USA, August 2009.

\bibitem{MS12}
Guillaume Mijoule, Nicolas Savy, and St\'ephanie Savy.
\newblock Models for patients recruitment in clinical trials and sensitivity
  analysis.
\newblock {\em Statistics in Medicine}, 31(16):1655--1674, 2012.

\bibitem{A2020}
Vladimir~V. Anisimov.
\newblock Modern analytic techniques for predictive modeling of clinical trial
  operations.
\newblock In Marchenko O.V. and Katenka N.V., editors, {\em Quantitative
  Methods in Pharmaceutical Research and Development: Concepts and
  Applications}, pages 361--408. Springer, 2020.

\bibitem{an-comm-stat11}
Vladimir~V. Anisimov.
\newblock Statistical modeling of clinical trials (recruitment and
  randomization).
\newblock {\em Comm. Statist. Theory Methods}, 40(19-20):3684--3699, 2011.

\bibitem{GSC08}
Byron~J. Gajewski, Stephen~D. Simon, and Susan~E. Carlson.
\newblock Predicting accrual in clinical trials with {B}ayesian posterior
  predictive distributions.
\newblock {\em Statistics in Medicine}, 27(13):2328--2340, 2008.

\bibitem{pmid20604946}
K.~D. Barnard, L.~Dent, and A.~Cook.
\newblock {{A} systematic review of models to predict recruitment to
  multicentre clinical trials}.
\newblock {\em BMC Med Res Methodol}, 10:63, Jul 2010.

\bibitem{pmid26188165}
D.~F. Heitjan, Z.~Ge, and G.~S. Ying.
\newblock {{R}eal-time prediction of clinical trial enrollment and event
  counts: {A} review}.
\newblock {\em Contemp Clin Trials}, 45(Pt A):26--33, Nov 2015.

\bibitem{pmid31299358}
E.~Gkioni, R.~Rius, S.~Dodd, and C.~Gamble.
\newblock {{A} systematic review describes models for recruitment prediction at
  the design stage of a clinical trial}.
\newblock {\em J Clin Epidemiol}, 115:141--149, 11 2019.

\bibitem{an-CCT-16}
Vladimir~V. Anisimov.
\newblock Discussion on the paper "{R}eal-time prediction of clinical trial
  enrollment and event counts: a review", by {DF} {H}eitjan et al.
\newblock {\em Contemporary Clinical Trials}, 46:7--10, 2016.

\bibitem{CONSORT}
D.~Moher, K.~F. Schulz, and D.~G. Altman.
\newblock {{T}he {C}{O}{N}{S}{O}{R}{T} statement: revised recommendations for
  improving the quality of reports of parallel-group randomised trials}.
\newblock {\em Lancet}, 357(9263):1191--1194, Apr 2001.

\bibitem{an-world-congr08}
Vladimir~V. Anisimov.
\newblock Using mixed {P}oisson models in patient recruitment in multicentre
  clinical trials.
\newblock In {\em Proceegings of the World Congress on Ingineering}, volume~II,
  pages 1046--1049, London, United Kingdom, 2008.

\bibitem{john93}
Norman~L. Johnson, Samuel Kotz, and Adrienne~W. Kemp.
\newblock {\em Univariate discrete distributions}.
\newblock Wiley Series in Probability and Mathematical Statistics: Applied
  Probability and Statistics. John Wiley \& Sons Inc., New York, second
  edition, 1992.
\newblock A Wiley-Interscience Publication.

\bibitem{bern04}
J.~M. Bernardo and A.~F.~M. Smith.
\newblock {\em Bayesian Theory}.
\newblock John Wiley \& Sons, Hoboken, NJ, USA, 2004.

\bibitem{Minois1}
N.~Minois, S.~Savy, V.~Lauwers-Cances, S.~Andrieu, and N.~Savy.
\newblock {{H}ow to deal with the {P}oisson-gamma model to forecast patients'
  recruitment in clinical trials when there are pauses in recruitment dynamic?}
\newblock {\em Contemp Clin Trials Commun}, 5:144--152, Mar 2017.

\bibitem{Minois2}
N.~Minois, V.~Lauwers-Cances, S.~Savy, M.~Attal, S.~Andrieu, V.~Anisimov, and
  N.~Savy.
\newblock {{U}sing {P}oisson-gamma model to evaluate the duration of
  recruitment process when historical trials are available}.
\newblock {\em Stat Med}, 36(23):3605--3620, Oct 2017.

\end{thebibliography}

\end{document}